\documentclass[pra,twocolumn,showpacs,preprintnumbers,superscriptaddress,amsmath,amssymb]{revtex4}
\usepackage{amsmath}

\usepackage{graphicx}
\usepackage{dcolumn}
\usepackage{bm}
\usepackage[dvipdfm,pdfstartview=FitH]{hyperref}

\begin{document}

\title{Transfer and teleportation of quantum states encoded in decoherence-free
subspace}

\author{Hua Wei}
\email{huawei.hw@gmail.com}

\affiliation{State Key Laboratory of Magnetic Resonance and Atomic
and Molecular Physics, Wuhan Institute of Physics and Mathematics,
Chinese Academy of Sciences, Wuhan, 430071, China}

\affiliation{Graduate School of the Chinese Academy of Sciences,
Beijing, 100049, China}

\author{ZhiJiao Deng}

\affiliation{State Key Laboratory of Magnetic Resonance and Atomic
and Molecular Physics, Wuhan Institute of Physics and Mathematics,
Chinese Academy of Sciences, Wuhan, 430071, China}
\affiliation{Graduate School of the Chinese Academy of Sciences,
Beijing, 100049, China}

\author{XiaoLong Zhang}

\affiliation{State Key Laboratory of Magnetic Resonance and Atomic
and Molecular Physics, Wuhan Institute of Physics and Mathematics,
Chinese Academy of Sciences, Wuhan, 430071, China}
\affiliation{Graduate School of the Chinese Academy of Sciences,
Beijing, 100049, China}

\author{Mang Feng}
\email{mangfeng1968@yahoo.com}

\affiliation{State Key Laboratory of Magnetic Resonance and Atomic
and Molecular Physics, Wuhan Institute of Physics and Mathematics,
Chinese Academy of Sciences, Wuhan, 430071, China}

\begin{abstract}
Quantum state transfer and teleportation, with qubits encoded in
internal states of the atoms in cavities, among spatially separated
nodes of a quantum network in decoherence-free subspace are
proposed, based on a cavity-assisted interaction by single-photon
pulses. We show in details the implementation of a logic-qubit
Hadamard gate and a two-logic-qubit conditional gate, and discuss
the experimental feasibility of our scheme.
\end{abstract}
 \maketitle

Quantum state transfer and teleportation are significant components
in quantum information processing, especially for quantum network.
As the confined atoms in cavity QED system are well suited for
storing qubits in long-lived internal states, spatially separated
cavities could be used to build a quantum network assisted by
photons \cite{cirac,photon,cpf,xue,ng,tr,deng,J}. On the other hand,
decoherence due to the inevitable interaction with environment
destroys quantum coherence. So decoherence-free subspaces (DFSs) of
Hilbert space has been introduced to protect against some errors due
to environmental coupling with certain symmetry
\cite{dfs0,dfs1,dfs2}. For example,  Ref. \cite{dfs2} utilized two
atoms to encode  single-logic-qubit, i.e.,
$|\widetilde{0}\rangle\equiv|1\rangle_1|0\rangle_2=|10\rangle$,
$|\widetilde{1}\rangle\equiv|0\rangle_1|1\rangle_2=|01\rangle$,
which are robust to collective dephasing error caused by ambient
magnetic fluctuation.

In this Brief Report, for the quantum state encoded in DFS mentioned
above, we present implementation of single-logic-qubit Hadamard gate
and two-logic-qubit conditional gate based on cavity-assisted
interaction with single-photon pulses. Based on these gates, we will
carry out the quantum state transfer and teleportation between two
spatially separated nodes in a quantum network. Compared with
previous related works, our proposal does not rely on the
synchronous optical lattices \cite{xue} in implementation of the
single-logic-qubit Hadamard gate. In addition, auxiliary entangled
photon pairs, as employed in \cite{ng}, are unnecessary in our
two-logic-qubit conditional gate. Moreover, for quantum state
transfer, neither the entangled photon pairs \cite{tr} nor the
special time-symmetric wave packet of the photons \cite{cirac} is
necessary in our scheme. So our scheme could not only protect
quantum information from some decoherence, but also reduce the
experimental difficulty compared to the previous schemes \cite
{cirac,xue,ng,tr}. Furthermore, in our scheme, each node of the
quantum network in DFS has individual input port for photons to
complete necessary operations, and different operational results can
be distinguished by detecting output photons.

\begin{figure}
\includegraphics[width=6.0cm]{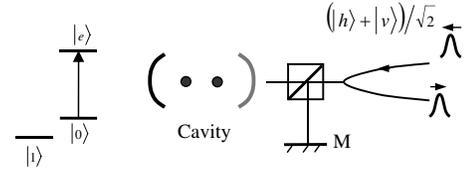}
\caption{\label{0}Left: Level configuration of an atom in a cavity.
Right: Schematic setup to implement a two-atom controlled phase gate
by cavity-assisted polarized-photon scattering.}
\end{figure}

The main idea using cavity-assisted photon scattering to realize a
controlled phase flip (CPF) between two atoms
\cite{cpf,xue,ng,tr,deng}, is sketched below. Suppose that two
identical atoms, each of which has a three-level configuration, are
well located in a high-finesse cavity. The levels $|0\rangle$ and
$|e\rangle$ of the atom are resonantly coupled by the bare cavity
mode with $h$ polarization or by the $h$ component of an input
photon, while level $|1\rangle$ is decoupled because of the large
detuning, as shown in Fig.~\ref{0}. If the duration of the input
photon pulse $T$ is sufficiently long and the atom-cavity coupling
is much stronger than both the cavity decay and spontaneous emission
of the atomic state, the pulse in resonance with the bare cavity
mode, i.e., the two atoms in the state $|1\rangle_1|1\rangle_2$,
would yield the pulse shape almost unchanged but with a $\pi$ phase
added when reflected by the cavity. On the contrary, if any of the
two atoms is in the state $|0\rangle$, as the cavity mode is shifted
by the resonance with the atom, the pulse will be reflected by the
cavity with both its shape and phase unchanged \cite{cpf}.
Therefore, by reflecting a single-photon pulse with $h$
polarization, the two-atom controlled phase gate $U_{\text{CPF}}
=\exp(i\pi|1\rangle_{11}\langle1|\otimes|1\rangle_{22}\langle1|)$ is
available.

\begin{figure}
\includegraphics[width=8.0cm]{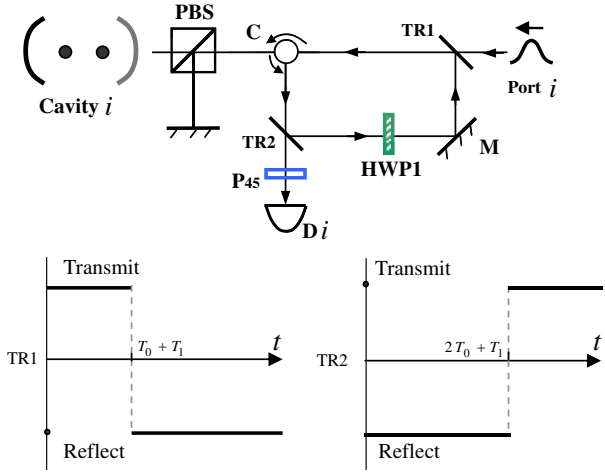}
\caption{\label{1}(Color online) Schematic setup for a single-
logic-qubit $\widetilde{H}$ operation and time control sequence of
TR1 and TR2. D${i}$ is a detector, C is a circulator. HWP1 rotates
the photon polarization as $|h\rangle\leftrightarrow|v\rangle$. TR
can be controlled exactly as needed to transmit or to reflect a
photon. The single-photon pulse's path for $\widetilde{H}$ is port
$i$-TR1-C-PBS-Cavity-PBS-C-TR2-HWP1-M-TR1-C-PBS-Cavity-PBS-C-TR2-P$_{45}$-D$i$.
The states of TR1 and TR2 can be controlled accurately by computer.
$T_0$ is the time for single-photon pulse process
TR1-C-PBS-Cavity-PBS-C-TR2, while $T_1$ is the time for
TR2-HWP1-M-TR1.}
\end{figure}

To carry out a quantum transfer and a teleportation in the DFS, we
need to construct a Hadamard operation $\widetilde{H}$ on the
single-logic qubit and a conditional gate operation on the two-logic
qubit. The $\widetilde{H}$ operation is shown in Fig.~\ref{1} for
the logic-qubit at arbitrary node $i$ (or, say, cavity $i$). The
half-wave plate 1 (HWP1), with its axis at $45^\circ$ to the
horizontal direction, rotates the photon polarization as
$|h\rangle\leftrightarrow|v\rangle$. P$_{45}$ is a $45^\circ$
polarizer projecting the polarization
$(|h\rangle+|v\rangle)/\sqrt{2}$. TR is a device which can be
controlled exactly as needed to transmit or reflect a photon
\cite{tr}, and the switching-time sequences of TR1 and TR2 are also
given in Fig.~\ref{1}.

Assume that the state of the logic-qubit in cavity $i$ is
$|\widetilde{1}\rangle$. The $\widetilde{H}$ operation can be
performed by following four steps. (1) A single-photon pulse in
state $(1/\sqrt{2})(|h\rangle+|v\rangle)$ is imported from port $i$.
It passes through TR1 and C, and then reaches the polarizing
beamsplitter (PBS) and the cavity $i$. Meanwhile, we perform a flip
operation on the logic-qubit inside the cavity $i$, i.e.,
$\widetilde{X}=C_{\text{NOT}}^{2,1}H_2U_{\text{CPF}}H_2C_{\text{NOT}}^{2,1}$
whose function is to flip the logic-qubit
$|\widetilde{0}\rangle\leftrightarrow|\widetilde{1}\rangle$
\cite{deng}, and thereby the state becomes
$(1/\sqrt{2})(|h\rangle|\widetilde{0}\rangle+|v\rangle|\widetilde{1}\rangle)$
after the photon pulse comes back from the cavity $i$ and the PBS.
(2) Reflected by TR2, the photon pulse goes through HWP1, yielding
the state
$(1/\sqrt{2})(|v\rangle|\widetilde{0}\rangle+|h\rangle|\widetilde{1}\rangle)$.
(3) After being reflected by M and TR1, the single-photon pulse
arrives at PBS again. We need two NOT operations on atom 1, one
before and another after the photon pulse is reflected by the
cavity, i.e., the operation sequence
$\sigma_{x}^1U_{\text{CFP}}\sigma_{x}^1$. So we get to the state
$(1/\sqrt{2})(|v\rangle|\widetilde{0}\rangle-|h\rangle|\widetilde{1}\rangle)$.
(4) Finally, when the single-photon pulse passes TR2 and P$_{45}$,
the photon is detected by D$i$, and thereby the logic-qubit inside
the cavity $i$ is left in the state
$(1/\sqrt{2})(|\widetilde{0}\rangle-|\widetilde{1}\rangle)$. The
operation process above can be shown more specifically as follows,
\begin{equation}
\begin{split}
\frac{|h\rangle+|\nu\rangle}{\sqrt{2}}|\widetilde{1}\rangle
\xrightarrow{\widetilde{X}}\frac{1}{\sqrt{2}}(|h\rangle|\widetilde{0}\rangle
+|\nu\rangle|\widetilde{1}\rangle)\\
\xrightarrow{\text{HWP1}}\frac{1}{\sqrt{2}}(|\nu\rangle|\widetilde{0}\rangle+|h\rangle|\widetilde{1}\rangle)
\xrightarrow{\sigma_{x}^1U_{\text{CFP}}\sigma_{x}^1}\\
\frac{1}{\sqrt{2}}(|\nu\rangle|\widetilde{0}\rangle-|h\rangle|\widetilde{1}\rangle)
\xrightarrow{\text{P}_{45}}\frac{|h\rangle+|\nu\rangle}{\sqrt{2}}\frac{1}{\sqrt{2}}(|\widetilde{0}\rangle-|\widetilde{1}\rangle).
\end{split}
\end{equation}
So a click of D$i$ means the success of the Hadamard gate on the
logic qubit. Otherwise, we have to repeat above steps from the very
beginning with the single photon input and initial atomic state
preparation. It is easy to check that Fig.~\ref{0} also works for
$|\widetilde{0}\rangle$ to be
$\frac{1}{\sqrt{2}}(|\widetilde{0}\rangle+|\widetilde{1}\rangle)$.
Compared with the previous operation of a single-logical-qubit
Hadamard gate \cite{xue}, our proposal needs neither the ancilla
system to store information nor the optical lattices to transfer
atoms synchronously across the cavity.

To ensure the success of above operations, we require the switching
time sequences of TR to be implemented accurately \cite{tr}.
Actually, in addition to the time control sequences of TR1 and TR2
designed in Fig. 2, we have an alternative. Take TR1 for instance:
Once the single-photon pulse from port $i$ has gone through TR1, we
change the transmitting state of the TR1 into reflecting state
without waiting for the time $T_0+T_1$. The switching-time sequence
designed for $\widetilde{H}$ is general for arbitrary states of the
single-logic qubit.

\begin{figure}
\includegraphics[width=8.0cm]{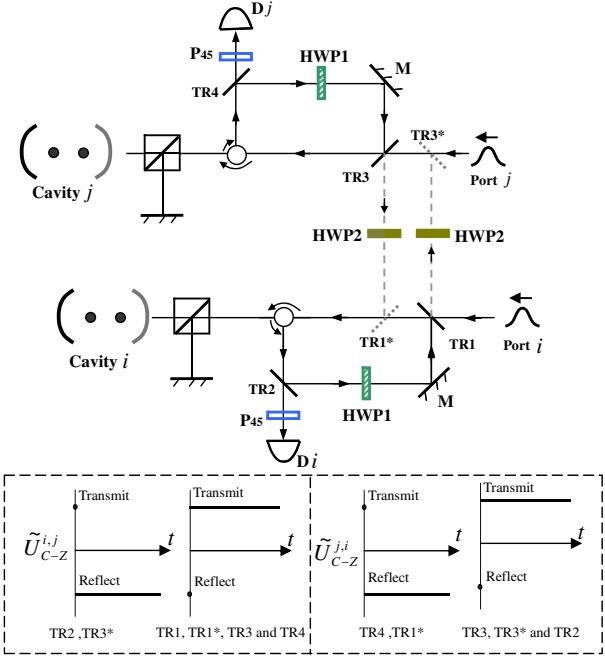}
\caption{\label{2} (Color online) Schematic setup for two
logic-qubit conditional phase gate $\widetilde{U}_{C-Z}^{i,j}$
($\widetilde{U}_{C-Z}^{j,i}$) which uses the additional channel
shown by dashed gray lines. HWP2 performs a Hadamard gate on the
photon polarization states. The single-photon pulse's path for
$\widetilde{U}_{C-Z}^{i,j}$ ($\widetilde{U}_{C-Z}^{j,i}$) is: port
$i$-TR1-TR1*-C-PBS-Cavity $i$-PBS-C-TR2-HWP1-M-TR1-HWP2-TR3*-TR3
-C-PBS-Cavity $j$-PBS-C-TR4-P$_{45}$-D$j$ (port
$j$-TR3*-TR3-C-PBS-Cavity $j$-PBS-C-TR4-HWP1-M-TR3-HWP2-TR1*-C-
PBS-Cavity $i$-PBS-C-TR2-P$_{45}$-D$i$). Ulike $\widetilde{H}$, the
TRs remain unchanged during the conditional phase gating.}
\end{figure}

A two nodes controlled phase gate for logic-qubits encoded in a DFS,
without using the entangle photon pairs as in \cite{ng}, can be
implemented with the aid of an additional channel shown by dashed
gray lines in Fig.~\ref{2}. HWP2, at an angle of $22.5^\circ$ to the
horizontal direction, performs a Hadamard gate on the photon
polarization states, i.e.,
$|h\rangle\rightarrow(1/\sqrt{2})(|h\rangle+|v\rangle)$,
$|v\rangle\rightarrow(1/\sqrt{2})(|h\rangle-|v\rangle)$.
$\widetilde{U}_{\text{C-Z}}^{i,j}$ is a controlled z (C-Z) phase-
flip gate with subscripts $i$ and $j$ indicating the control and
target logic-qubits in cavities $i$ and $j$, respectively. Assume
that the two logic-qubits are in superposition state,
$a|\widetilde{0}\rangle_i|\widetilde{0}\rangle_j
+b|\widetilde{0}\rangle_i|\widetilde{1}\rangle_j+
c|\widetilde{1}\rangle_i|\widetilde{0}\rangle_j+
d|\widetilde{1}\rangle_i|\widetilde{1}\rangle_j$, where $a,b,c$ and
$d$ are normalized constants. The $\widetilde{U}_{\text{C-Z}}^{i,j}$
operation is completed by following three steps. (1) A single-photon
pulse in state $(1/\sqrt{2})(|h\rangle+|v\rangle)$ is input from the
port $i$, passes through PBS, and reaches the cavity $i$. We need to
perform following operations
$\sigma_{x,i}^2U_{\text{CPF}}\sigma_{x,i}^2$, i.e., $\sigma_x$
operation on the atom 2 in cavity $i$ before and after the
single-photon pulse is reflected by the cavity $i$. (2) Due to
reflection of TR2, the single-photon pulse passes through HWP1 and
goes into the additional channel connected to the node $j$. (3)
After going through HWP2 and reflected by TR3*, the single-photon
pulse arrives at the PBS and then the cavity $j$. As previously, we
have to perform $\sigma_{x,j}^1U_{\text{CPF}}\sigma_{x,j}^1$, i.e.,
the $\sigma_x$ operation, on the atom 1 in the cavity $j$ before and
after the single-photon pulse is reflected by the cavity $j$. Then
the single-photon pulse goes through $\text{P}_{45}$ and would be
detected by D$j$. The $\widetilde{U}_{\text{C-Z}}^{i,j}$ operation
process above can be shown step by step as
\begin{equation}
\begin{split}
&\frac{1}{\sqrt{2}}(|h\rangle+|v\rangle)\otimes(a|\widetilde{0}\rangle_i|\widetilde{0}\rangle_j
+b|\widetilde{0}\rangle_i|\widetilde{1}\rangle_j+c
|\widetilde{1}\rangle_i|\widetilde{0}\rangle_j+d
|\widetilde{1}\rangle_i|\widetilde{1}\rangle_j)\\
&\xrightarrow{\sigma_{x,i}^2U_{\text{CFP}}\sigma_{x,i}^2}
\frac{1}{\sqrt{2}}[a(-|h\rangle+|v\rangle)|\widetilde{0}\rangle_i|\widetilde{0}\rangle_j\\
&+b(-|h\rangle+|v\rangle)|\widetilde{0}\rangle_i|\widetilde{1}\rangle_j+
c(|h\rangle+|v\rangle)|\widetilde{1}\rangle_i|\widetilde{0}\rangle_j\\
&+d(|h\rangle+|v\rangle)|\widetilde{1}\rangle_i|\widetilde{1}\rangle_j]
\xrightarrow{\text{HWP1,HWP2}}a|v\rangle|\widetilde{0}\rangle_i|\widetilde{0}\rangle_j\\
&+b|v\rangle|\widetilde{0}\rangle_i|\widetilde{1}\rangle_j+
c|h\rangle|\widetilde{1}\rangle_i|\widetilde{0}\rangle_j+
d|h\rangle|\widetilde{1}\rangle_i|\widetilde{1}\rangle_j\\
&\xrightarrow{\sigma_{x,j}^1U_{\text{CFP}}\sigma_{x,j}^1,\text{P}_{45}}
\frac{1}{\sqrt{2}}(|h\rangle+|v\rangle)\otimes(a|\widetilde{0}\rangle_i|\widetilde{0}\rangle_j\\
&+b|\widetilde{0}\rangle_i|\widetilde{1}\rangle_j+
c|\widetilde{1}\rangle_i|\widetilde{0}\rangle_j-
d|\widetilde{1}\rangle_i|\widetilde{1}\rangle_j).
\end{split}\label{cz}
\end{equation}
If we get a click of D$j$, the $\widetilde{U}_{\text{C-Z}}^{i,j}$
operation succeeds. The silence of D$j$ means failure and we must
repeat above steps from the very beginning. Considering the
symmetric operation for the two qubits gate, the
$\widetilde{U}_{\text{C-Z}}^{j,i}$ operation can be done by a
similar process to Eqs.(\ref{cz}), where the single-photon pulse is
input from port $j$, as shown in Fig.~\ref{2}. The corresponding
controlled-NOT(CNOT) operation for the logic qubits in a DFS is
given by $\widetilde{C}^{i,j}_{\text{NOT}}=\widetilde{H}_j\otimes
\widetilde{U}_{\text{C-Z}}^{i,j}\otimes\widetilde{H}_j$ and
$\widetilde{C}^{j,i}_{\text{NOT}}=\widetilde{H}_i\otimes
\widetilde{U}_{\text{C-Z}}^{j,i}\otimes\widetilde{H}_i$. It should
be mentioned that, the TR1* (or TR3* for node $j$), whose function
is to connect spatially separated nodes, has no action during the
single-logic-qubit Hadamard operation. So we can keep them always on
in transmitting state during the time the $\widetilde{H}$ operation
is carried out in individual nodes.

The transfer of information in a quantum network is an important
subject for quantum information processing \cite{cirac,photon}.
Based on the operations investigated above, we realize below the
information exchange between the $i$th logic-qubit
$\alpha_i|\widetilde{0}\rangle+\beta_i|\widetilde{1}\rangle$ and the
$j$th logic-qubit
$\alpha_j|\widetilde{0}\rangle+\beta_j|\widetilde{1}\rangle$, using
following single-photon pulse sequence:
\begin{equation}
\widetilde{U}_{\text{SWAP}}^{i,j}=\widetilde{C}^{i,j}_{\text{NOT}}
\widetilde{C}^{j,i}_{\text{NOT}}\widetilde{C}^{i,j}_{\text{NOT}}.
\end{equation}
The corresponding process reads
$(\alpha_i|\widetilde{0}\rangle+\beta_i|\widetilde{1}\rangle)
(\alpha_j|\widetilde{0}\rangle+\beta_j|\widetilde{1}\rangle)
\rightarrow(\alpha_j|\widetilde{0}\rangle+\beta_j|\widetilde{1}\rangle)
(\alpha_i|\widetilde{0}\rangle+\beta_i|\widetilde{1}\rangle)$.
Compared with previous work \cite{cirac}, we do not need the special
laser pulses with time-dependent Rabi frequency and laser phases to
excite a time-symmetric wave packet of the photon from the sending
node to the receiving node. In addition, we do not require a
transfer channel made using auxiliary entangled photon pairs
\cite{tr}. More importantly, compared to \cite{cirac,tr}, our
logic-qubits are encoded in DFS which is robust to collective
dephasing error.

\begin{table}
\caption{\label{t} Bob's corresponding operations on the logic-qubit
$|\widetilde{\Psi}\rangle_k$ after he has learned the measurement
outcomes from Alice by a classical communication channel.
$\sigma_{x,k}^1$ means $\sigma_x$ operation on the atom 1 in cavity
$k$. The $\widetilde{X}$ operation has been given in the text.}
\begin{ruledtabular}
\begin{tabular}{ccc}
Alice's measurements & Bob's operations\\
\hline\\
$|\widetilde{0}\rangle_i|\widetilde{0}\rangle_j$&Nothing\\
$|\widetilde{0}\rangle_i|\widetilde{1}\rangle_j$&$\widetilde{X}$\\
$|\widetilde{1}\rangle_i|\widetilde{0}\rangle_j$&$\widetilde{Z}=\sigma_{x,k}^1U_{\text{CPF}}\sigma_{x,k}^1$\\
$|\widetilde{1}\rangle_i|\widetilde{1}\rangle_j$&$\widetilde{Z}\widetilde{X}$
\end{tabular}
\end{ruledtabular}
\end{table}

Now we turn to a proposal for quantum state teleportation
\cite{bennett} for the logic qubits described above. Assume that we
have three logic-qubits in states $|\widetilde{\Psi}\rangle_i$,
$|\widetilde{\Psi}\rangle_j$ and $|\widetilde{\Psi}\rangle_k$,
respectively, with the former two belonging to Alice and the third
to Bob.
$|\widetilde{\Psi}\rangle_i=\alpha|\widetilde{0}\rangle+\beta|\widetilde{1}\rangle$
is an unknown state and would be teleported to Bob. Firstly,
teleportation need entanglement between Alice's second qubit
$|\widetilde{\Psi}\rangle_j$ and Bob's qubit
$|\widetilde{\Psi}\rangle_k$ in a Bell state. We take
$|\widetilde{\Psi}\widetilde{\Psi}\rangle_{jk}
=\frac{1}{\sqrt{2}}(|\widetilde{0}\rangle_j|\widetilde{0}\rangle_k
+|\widetilde{1}\rangle_j|\widetilde{1}\rangle_k)$ for example. The
Bell-style entanglement can be implemented by $\widetilde{H}$ and
$\widetilde{C}_{\text{NOT}}$ operation for the initial state
$|\widetilde{0}\rangle_j|\widetilde{0}\rangle_k$, i.e.,
\begin{equation}
\begin{split}
|\widetilde{0}\rangle_j|\widetilde{0}\rangle_k
\xrightarrow{\widetilde{H}_j}\frac{1}{\sqrt{2}}(|\widetilde{0}\rangle_j+|\widetilde{1}\rangle_j)
|\widetilde{0}\rangle_k\\
\xrightarrow{\widetilde{C}_{\text{NOT}}^{j,k}}
\frac{1}{\sqrt{2}}(|\widetilde{0}\rangle_j|\widetilde{0}\rangle_k
+|\widetilde{1}\rangle_j|\widetilde{1}\rangle_k).
\end{split}
\end{equation}
The Bell state measurement in teleportation also uses
$\widetilde{H}$ and $\widetilde{C}_{\text{NOT}}$ operations, i.e., a
$\widetilde{C}_{\text{NOT}}^{i,j}$ operation between
$|\widetilde{\Psi}\rangle_i$ and $|\widetilde{\Psi}\rangle_j$ and an
$\widetilde{H}_i$ operation to $|\widetilde{\Psi}\rangle_i$. As a
result, we get the final state
$|\widetilde{\Phi}\rangle=\frac{1}{2}[|\widetilde{0}\rangle_i|\widetilde{0}\rangle_j
(\alpha|\widetilde{0}\rangle+\beta|\widetilde{1}\rangle)_k+|\widetilde{0}\rangle_i|\widetilde{1}\rangle_j
(\alpha|\widetilde{1}\rangle+\beta|\widetilde{0}\rangle)_k+|\widetilde{1}\rangle_i|\widetilde{0}\rangle_j
(\alpha|\widetilde{0}\rangle-\beta|\widetilde{1}\rangle)_k+|\widetilde{1}\rangle_i|\widetilde{1}\rangle_j
(\alpha|\widetilde{1}\rangle-\beta|\widetilde{0}\rangle)_k]$.
Following the operations shown in Table \ref{t}, we finish the
teleportation of a quantum state from the logic-qubit $i$ to the
logic-qubit $k$.

From the schematic setup in Fig.~\ref{2} and the operations
presented above, we can learn that the single-logical-qubit Hadamard
operation and the two-logical-qubit conditional operation are
coexisting in our scheme and could work independently by controlling
the transmitting or reflecting state of the connecting devices TR1*
and TR3*. This is very important for scalability of the quantum
network.

Experimentally, the currently achieved technology of deterministic
single-photon source \cite{hij} provides potential support for our
scheme. As 300,000 high-quality single photons could be generated
continuously within 30 sec, a fast implementation of our scheme is
available. Moreover, to fix two atoms in an optical cavity, we have
to confine the atoms in optical lattices embedded in an optical
cavity, and this has been achieved experimentally \cite {sauer}. To
confine each atom in a particular lattice, a more advanced technique
is needed. Alternatively, we may consider an ion-trap-cavity
combinatory setup with two charged atoms fixed by the trap potential
and optically coupled by the cavity mode. A single Calcium ion has
been successfully trapped in such a device \cite {blatt}. The
experimental extension to two ions satisfying Purcell condition
would be in principle available in the near future.

In our scheme, if the duration $T$ for the photon pulse input in the
cavity and the cavity decay rate $\kappa$ satisfies $\kappa T\gg 1$,
the basic operation $U_{\text{CPF}}$ is insensitive to both the
atom-cavity coupling strength and the Lamb-Dicke localization.
According to the numerical simulation \cite{cpf}, if $\kappa
T\sim100$ and the atom-cavity coupling is several times stronger
than the dissipative factors of the system, the gate fidelity is
almost unity. In terms of the numerical \cite{cpf} and the
experimental results \cite{J}, a $U_{\text{CPF}}$ takes $T=3\sim5\mu
s$ for $\kappa T\gg1$. With the experimental numbers
$\kappa/2\pi\sim4$ MHz, $g/2\pi\sim30$ MHz, $\Gamma/2\pi\sim2.6$ MHz
\cite{J,time}, we may estimate the time for the logic-qubit
operations $\widetilde{H}$ and $\widetilde{C}_{\text{NOT}}$ to be of
the order of $10^{-6}\sim10^{-5}s$. In realistic experiments,
however, we should pay attention to photon loss during the
experiment and to the detector efficiency. Fortunately, as the
different operational results in each step are distinguishable in
our scheme, (for example, the single-logic-qubit operation
$\widetilde{H}_i$ in the $i$th cavity is associated with a
single-photon pulse input in port $i$ and the detector D$i$, whereas
the two-logic-qubit operation $\widetilde{U}_{\text{C-Z}}^{i,j}$ is
relevant to the input port $i$ and the response of D$j$), the
successful detections of the single-photon pulses ensure the high
fidelities of the gate operations. That implies that, as we discard
the events without detector clicks, the errors due to photon loss
can be completely removed, i.e., this is a repeat-until-success
operations \cite{repeat}. So, for our proposal, high-fidelity
quantum gatings are possible even in the case of the photon loss
\cite{p}.

In summary, we have proposed the implementation of a
single-logic-qubit Hadamard operation and a two-logic-qubit
conditional gate, based on which we may carry out quantum state
transfer and teleportation between spatially separated nodes of a
quantum network in a DFS. As the cavity QED technique develops very
quickly, we hope for more applications of the quantum logic
operation in DFS regarding our scheme.

This work is partly supported by National Natural Science Foundation
of China under Grant Nos. 10474118 and 60490280, by Hubei Provincial
Funding for Distinguished Young Scholars, and partly by the National
Fundamental Research Program of China under Grant No. 2005CB724502
and No. 2006CB921203.

\end{document}